# The correlation between molecular motions and heat capacity in normal ice and water

Hai Bo Ke, Ping Wen[*], Wei Hua Wang

Institute of Physics, Chinese Academy of Sciences, Beijing 100190, P. R. China

Abstract

The heat capacities of ice and water at ambient pressure are reexamined to build an intrinsic correlation between $H_2O$ molecular motions and the heat capacity. Based on the evolution of $H_2O$ molecular motions, a satisfactory description of the heat capacity of ice and water is provided. The heat capacity of ice is related not only to $H_2O$ molecular vibrations, but also to the molecular rotations. In water, all $H_2O$ molecular vibrations, rotations and translations contribute to the heat capacity. The molecular translational motions are found to be the main contribution to the large heat capacity of water. The results provide a deep insight into the nature of water and ice at ambient pressure.

Heat capacity $C_P$ at constant pressure or $C_V$ at constant volume is one of fundamental properties of a matter. Based on the accurate description of the $C_P$ and $C_V$ in ideal solids and gases, a deep understanding to the nature of real solids and gases has been given [1, 2]. The $C_P$ and $C_V$ have a close relation with the intrinsic motions. For the ideal gas containing molecules, the $C_P$ is equal to $C_V+R$ ($R$ is gas constant). The $C_V$ is a function of the number of degrees of freedom of translational, rotational and vibrational motions [2]. The appearance for given type of motion is depended on temperature and pressure. For the ideal solid (Einstein solid [1]), $C_P$, as well as $C_V$, is close to the $C_V^{Vib}$ contributed from vibrations within a small error. The description of $C_V^{Vib}$, proposed firstly by Einstein [3] and then modified by Debye [4], compares well with the rule of Dulong and Petit [5]. Then, the vibration around equilibrium site has been considered as the basic characteristic of solids, even though the $C_P$ or $C_V$ of most solids containing molecules can not be described well with only the consideration of vibrations [6-8]. Hitherto, no ideal model for liquids has been founded yet. An accurate description of the $C_P$ and $C_V$ for liquids is still far away from our understanding. As one of the most important condensed matters in our planet, ice or water has the known compositional unit, $H_2O$ molecule. The molecule can be considered as a rigid bent because the thermal energy can not activate the atomic vibration inside the molecule at ambient pressure in the temperature region from 0 to 373 K. So, ice or water, a pretty simple condensed matter system formed by molecules, offers an opportunity to understand the origin of the heat capacity in a real condensed matter. Unfortunately, to describe the heat capacity $C_P$ of normal water and ice is still a problem unresolved, even though the $C_P$ has been measured for several decades [9-13].

In this report, we make an attempt to give a clear description to the $C_P$ for the water and ice on the basic of motions. The motion, carried by whole $H_2O$ molecule, involves molecular vibration, molecular rotation and molecular translation. It is found that the temperature dependence of the $C_P$ of ice corresponds to the evolution of rigid $H_2O$ molecular vibrations and rotations. No contribution of the translational motion exists in normal ice region from 0 to 273 K. The $C_P$ of water, almost independent of temperature from 273 to 373 K, originates from all of $H_2O$ molecular motions. The translational contribution is clarified by the exact ways to translate for $H_2O$ molecule in water. A simple model for water is founded, where the translational motion is its fundamental feature.

Figure 1 shows the heat capacity $C_P$ at ambient pressure for one molar ice and water. The complete $C_P$ picture in temperature region from 2 to 373 K is constructed with the previous experimental data [9, 10, 14]. It is found that the $C_P$ of ice increases gradually as temperature increases from 2 to the melting point, 273 K. At around 273 K, the $C_P$ is up to $\frac{9}{2}R$ ($R$ is gas constant). Remarkably, the $C_P$ of water is close to $\frac{18}{2}R$, and is almost unvaried with temperature in the region from 273 to 373 K. It is usual, in theoretical considerations, to ignore the difference between the $C_P$ and $C_V$ for a condensed matter; this neglect involves only small errors, and can be remedied if the accuracy of the theory should warrant it. The $C_P$ in Fig. 1 can be regarded as the $C_V$ for ice and water.

Firstly, the vibrational contribution to the $C_P$ in ice is focused. With the consideration that the atomic vibrations inside $H_2O$ molecule are neglected and one $H_2O$ molecule interacting with its neighbors only has three independent vibrations, we use Debye model to fit the experimental data of $C_P$ [as shown in Fig. 2(a)]. Different from the previous work where different sizes of vibrational unit was proposed[10], we make sure that ice containing $N$ $H_2O$ molecules has exactly $3N$ vibrations. The $C_P^{Vib}$ related to $H_2O$ molecular vibration in the ice has the form: $C_P^{Vib} \approx C_V^{Vib} = 9Nk_B \left(\frac{T}{\Theta_D(T)}\right)^3 \int_0^{\Theta_D(T)/T} \frac{x^4 e^x}{(e^x-1)^2} dx$ [4], where $\Theta_D(T)$ is temperature-dependent Debye temperature and $k_B$ is Boltzmann constant. The values of the $\Theta_D$ determined by the best fitting with the above equation are plotted in Fig. 2(b).

The tendency of the $\Theta_D$ with temperature is a little strange. The $\Theta_D$ decreases firstly to a minimum at around 13 K, and then increases to a maximum at about 86 K as temperature increases from 2 K. Especially, the maximum value of the $\Theta_D$ (319 K) is unreasonable because the melting point of the normal ice at ambient pressure is only 273 K. The possible reason is that this high value does not represent the tetrahedrally hydrogen bonded $H_2O$ molecules cluster that has been considered as the basic structure of ice [11], but the structure of "plane"-hexagonal $H_2O$ molecules array in ice. The interactions of $H_2O$ molecules in the array are stronger than those in the tetrahedrally hydrogen bonded cluster. Computer simulation [15] revealed that the "plane"-hexagonal $H_2O$ molecules array could exist at temperature up to 513 K while the tetrahedral hydrogen bonded cluster almost disappeared at 273 K.

The decrease of $\Theta_D$ at temperature above 86 K indicates that the $C_P$ of ice contains not only the virational contribution, but also other contributions. The similar phenomenon has been found in simple crystals, and explained by the appearance of anharmonicity [16-18]. Flubacher challenged it, and concluded the decrease was due to the excitation of the librational modes of vibration with the consideration of the $\Theta_D$ curves for $3N$ and $6N$ degrees of vibrational freedom [10]. However, the existence of librational modes of vibration is much doubtable since in theory no other modes of vibration can exist except the molecular vibration in ice. The estimated contribution of the translational modes [] is also unreasonable since the characteristic of rigid solid for ice is lost. We conclude that the other mode of $H_2O$ molecular motion appearing at high temperature must be $H_2O$ molecular rotation. This idea is supported by the Bernal-Fowler-Pauling ice rules [19, 20]. At high temperature rearrangements of $H_2O$ molecule in normal ice can take place only through the rotations of some $H_2O$ molecules with the position of oxygen atom fixed at a lattice [11, 21].

In order to deduce $C_V^{Rot}$ from the $C_P$ in Fig. 2(a), it is appropriate to flat the $\Theta_D$ off in the vicinity of 87 K [see Fig. 2(b)]. It is due to that the thermal motion of oxygen atoms in ice can be described by a single Debye characteristic temperature in the range from 183 to 273 K []. One can find that the value of $C_V^{Rot}$ is close to 0 at the temperature less than about 90 K. As temperature increases from 90 to 273 K, the $C_V^{Rot}$ increases monotonously from 0 to $\frac{3}{2}R$. The separation of the $C_P$ into $C_V^{Rot}$ and $C_P^{Vib}$ two parts not only confirms two types of molecular motions exist in normal ice at certain temperature, but also reveals the feature of molecular rotations in the ice. The tendency of $C_V^{Rot}$ indicates that to activate $H_2O$ molecular rotators in ice is a process. In theory, the rotation of one free $H_2O$ molecule has three degrees of freedom. Correspondingly, the heat capacity $C_V^{Rot}$ arisen from the rotation of one molar molecules is $\frac{3}{2}R$ when temperature is higher than $\frac{\hbar^2}{2\kappa_B I}$ ($\hbar$ is reduced Plank constant, $I$ is the moment of inertia of the molecule.) [2]. The value of $I$ for $H_2O$ molecule is higher than $1.0\times10^{-47}$ kg·m² [22]. Thus, the rotational contribution for one molar $H_2O$ molecule is $\frac{3}{2}R$ only when temperature is higher than 40 K. In fact, $H_2O$ molecules in ice are restricted by four nearest $H_2O$ molecules with hydrogen-bond. The temperature (corresponding to the thermal energy) needed to activate $H_2O$ molecule rotator must be higher than 40 K in order to overcome the energy barrier. Fig. 2(b) shows that the appearance of the molecular rotation is around 90 K. Moreover, the energy to the rotation is of a distribution corresponding to the heterogeneous strength and orientation of hydrogen-band in ice. Therefore, with the increasing temperature the number of $H_2O$ molecules to rotate increase. The corresponding $C_V^{Rot}$ in ice, direct proportion to the number of activated $H_2O$ rotator, increases gradually. At the temperature close to 273 K, all $H_2O$ molecules in ice have been activated to rotate freely, the value of the $C_V^{Rot}$ in ice is approaching to $\frac{3}{2}R$.

The $C_V^{Rot}$ in ice is also consistent with the residual entropy of ice [23]. Pauling has described the residual entropy of ice with $H_2O$ molecular rotation in the tetrahedral cluster [19]. $H_2O$ molecular rotators in ice are frozen out off the equilibrium randomly and gradually as temperature decrease. Finally, when temperature is less than 90 K all rotators can be frozen. The possible number of configurations for the ice containing $N$ frozen $H_2O$ molecular rotators will be $\left(\frac{6}{4}\right)^N$. The resultant residual entropy of ice containing one mol molecules is $R\ln\frac{3}{2}$, and compares very favorably with the experimental value of 3.43 Jmol$^{-1}$K$^{-1}$ [23]. Moreover, the origin of the $C_V^{Rot}$ in ice means that the suggestion of the structural transformation of pure hexagonal (normal) ice to an ordered phase at temperature less than 90 K [24] is not reasonable, even though the ordered phase is of a size in nanometers. The energy restricting H2O molecular rotation in the ordered phase is not of a distribution. The corresponding $C_V^{Rot}$ in ice must change dramatically at certain temperature. It is contrary to the result in Fig. 2(a).

$H_2O$ molecular motion in water must be clarified to understand the constant $C_P$ of water. Fortunately, the molecular motions in water have been known well in details [25-27]. In water, all types of motions involving in $H_2O$ molecular vibration, rotation and translations have been observed. $H_2O$ molecular vibrations in water characterized by the maximum vibrational amplitude perpendicular to the hydrogen bond can be deduced from the Debye-Waller factor. Time-resolved infrared spectroscopy has shown that the vibrational dynamics in water are dominated by underdamped displacement of hydrogen-bond coordinate on the very short time scale (<200 *fs*) [26, 27]. Besides the molecular vibration, two relaxations with different relaxation times have been clearly identified by the high-quality quasi-elastic incoherent neutron scattering [25]. The relaxation with short relaxation time is associated with the fluctuation of the hydrogen bond and related to the rotational diffusion of $H_2O$ molecules in water. At room temperature, its relaxation time is ~1 *ps*. At the intermediate-time scale (larger than 1 *ps*), a sufficient number of hydrogen bonds are broken, $H_2O$ molecule then jumps to the nearest site. The average distance of the jumps is about 1.6 Å. The motions have been well explained by the random jump/translational diffusion model [28]. Then, along the probing time a naïve picture to $H_2O$ molecular motions in water is given. In Fig. 3 $H_2O$ molecules in water are posited in a cage formed by their nearest neighbor $H_2O$ molecules when the probing time is on scales of hundreds of femtoseconds. In the cages, the vibration of $H_2O$ molecule takes place on the time scales of hundreds *fs*. As the probing time approaches to 1 *ps*, the rotation of $H_2O$ molecule appears. When the probing time is prolonging further, the jumps/translational motion of $H_2O$ molecule is observed. In a normal experiment such as the measurement of heat capacity on the time scales of several seconds, all types of motions including $H_2O$ molecular vibrations, rotations and translational motions exist in water. Correspondingly, in theoretical considerations the heat capacity of water contains three parts: $C_V^{Vib}$, $C_V^{Rot}$ and $C_V^{Tr}$.

It is clear that the values of the $C_V^{Vib}$ and $C_V^{Rot}$ in the water containing one mole $H_2O$ molecules are equal to $3R$ and $\frac{3}{2}R$ because all of $H_2O$ molecular vibrations and rotations in ice have been activated already at temperature around 273 K (see Fig. 2). The description of the $C_V^{Tr}$ related to molecular translation in its gaseous phase can not be applied directly into that in water. In gas, $H_2O$ molecular translational motions are complete independent since no interactions between $H_2O$ molecules exist. Its $C_V^{Tr}$ arisen from the translations of one mol $H_2O$ molecules is $\frac{3}{2}R$ since one $H_2O$ molecule has three degrees of translational freedom in space. In liquid, $H_2O$ molecular translations can not be described well without its rotations. Figure 3 displays the different ways to translate for the molecule. Prior to $H_2O$ molecular translational motion, the rotational motion has already taken place. Due to its rotation that is restricted by hydrogen-bond, one $H_2O$ molecule is transformed into three independent molecular forms with same possibility to translate in space. In Fig. 3 we image naively the three $H_2O$ molecular forms as three independent rotators along three independent orientations (x, y and z axis) in space. It is noted that here each independent rotator is defined to has one degree of rotational freedom, but two degrees. As the probing time is long enough, each molecule with a given molecular form can jump/ diffuse translationally and freely, and has three degrees of translational freedom. Then in water one $H_2O$ molecule with three degrees of rotational freedom will have 9 degrees of translational freedom (see Fig.3). Correspondingly, the $C_V^{Tr}$ for water containing one mol $H_2O$ molecules is equal to $\frac{9}{2}R$. In total, the specific heat of water is sum of the $C_V^{Vib}$, $C_V^{Rot}$ and $C_V^{Tr}$. Its value ($C_V = C_V^{Vib} + C_V^{Rot} + C_V^{Tr} = \frac{18}{2}R$) is exactly equal to the experimental data (see Fig. 1). This description to the heat capacity of water can be applied well into other liquids. But more attention must be paid since usually the motions inside molecules can not be neglected. The atomic vibrations and rotations of atomic group inside molecule give contributions to the heat capacity. For simply liquids like metallic liquids, there are no effects of atomic rotational motions on the way to translate. The heat capacity of metallic liquids containing one mole of atoms mainly contains two parts: the $C_V^{Vib}$ (3$R$) and $C_V^{Tr}$ ($\frac{3}{2}R$). The sum of them is 4.5$R$ (around 37 JK$^{-1}$mol$^{-1}$), and consistent with the fact that most pure liquid elements have $C_P$ values ranging between 30 and 40 JK$^{-1}$mol$^{-1}$ at the melting point [29].

The correlation of the heat capacity of normal ice and water with H$_2$O molecular motions displays a clear difference between the ice and water. The molecular translations are visible in water, while they are invisible in ice within the experimental time to measure the heat capacity. It is consistent with the basic characteristic of normal liquid that is well known [18]. A liquid can not support any shear within the experimental time scale. If one considers an atom or a molecule in the liquid state at any moment, it will be interacting with the neighbors, vibrating as though it were an atom or a molecule in the solid state. At the next moment with certain gap in time to the previous moment, the atom/molecule may already move away from its previous site. This translational motion must be activated thermally, and need overcome an energy barrier $\Delta E$. The measurement of the heat capacity might offer a clear understanding the intrinsic difference between liquid and rigid solid.

In summary, the correlation between intrinsic motions and the $C_P$ of normal water and ice is founded. In ice, translational motion of H$_2$O molecule is invisible at the probing time scale to measure the $C_P$, and only the vibrations and rotations of H$_2$O molecular have their contribution to the $C_P$. In water, all H$_2$O molecular vibrations, rotations and translations contribute to the $C_P$. The high value of the $C_P$ is mainly due to H$_2$O molecular translations. The degrees of the translational freedom per H$_2$O molecules are nine, and equal to three times of the degrees of rotational freedom per molecule in water. The findings might give a further insight into the difference between solids and liquids.

The work was supported by the Science Foundation of China (Grant Nrs: 51071170, 50731008, 50890171, and 50921091) and MOST 973 of China (No. 2007CB613904).

**Captions:**

Figure 1. The temperature dependence of the heat capacity $C_P$ for one mole of ice and water at ambient pressure. The data in Ref (*9, 10, 14*) marked by blue are plotted; the data marked by red are determined by differential scanning calorimeter (DSC) at the heating rate of 20 K/min, the wine curve represents the $C_P$ measured by DSC at the cooling rate of 20 K/min.

Figure 2. (a) The separation of the $C_P$ of ice into two parts: the $C_V^{Vib}$ arisen from H$_2$O molecular vibration and $C_V^{Rot}$ related to H$_2$O molecular rotation. The $C_V^{Vib}$ is calculated by Debye model with the Debye temperature $\Theta_D$ marked with red color in Fig.2 (b); the black line is the $C_V^{Rot}$ derived from $C_P - C_V^{Vib}$. (b) The temperature dependence of the $\Theta_D$ in ice.

Figure 3. The naïve scheme to the evolution of H$_2$O molecular motion along the probing time. The blue arrows denote the directions of H$_2$O molecular vibrations and translations. Noted that the number of the degrees of rotational motions along X, Y, and Z axis is 2, but here we define it as 1.

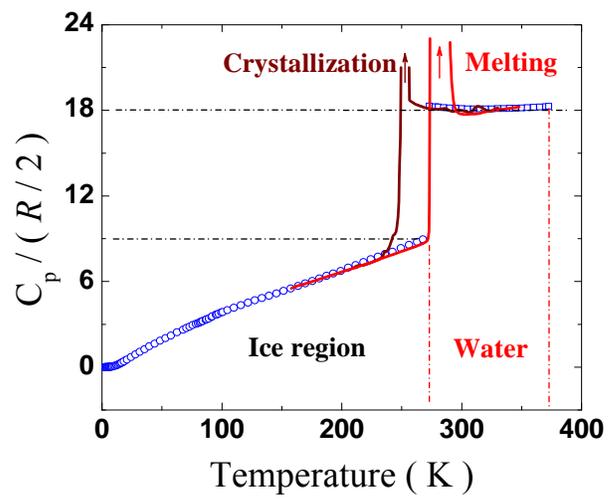

Figure 1. Ke, *et.al.,*

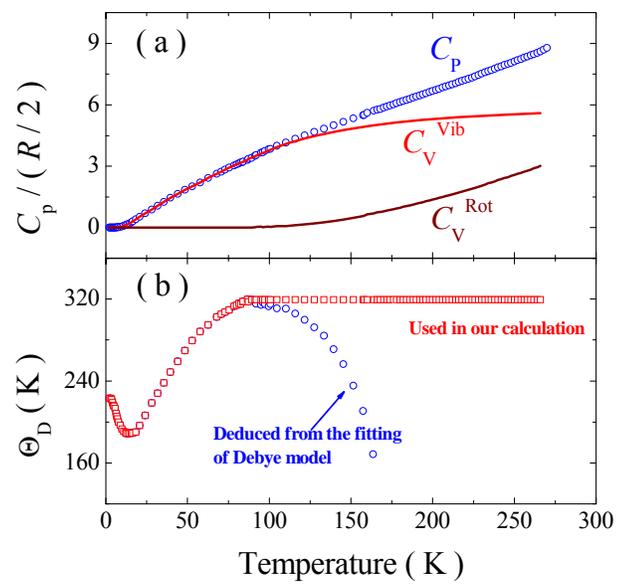

Figure 2. Ke, *et.al.,*

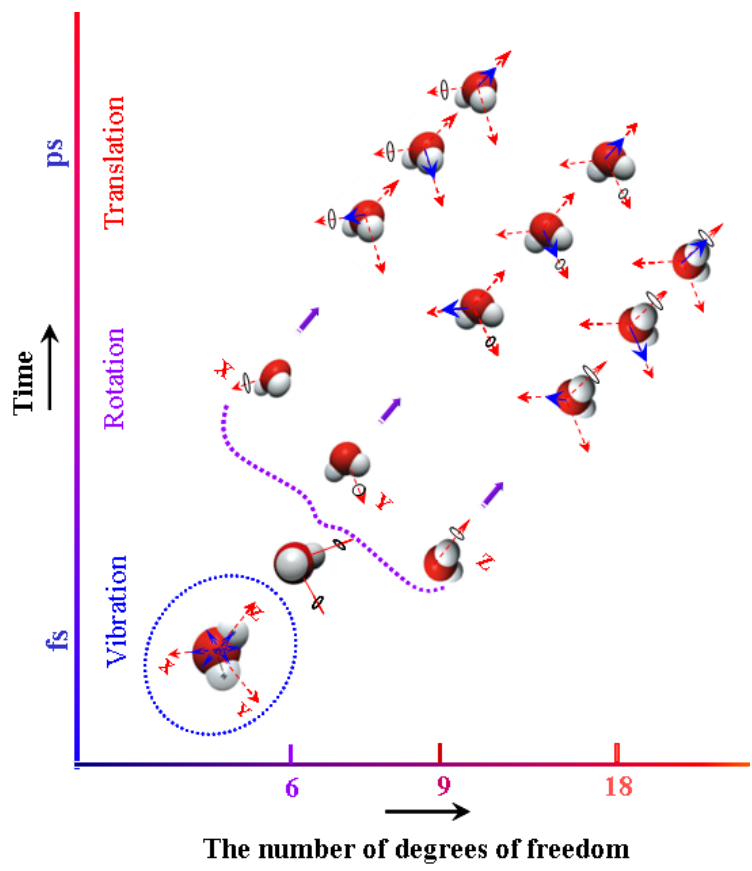

Figure 3. Ke, *et.al.,*